\documentclass[12pt]{article}
\usepackage{amsmath,amssymb}
\usepackage[dvipdfmx]{graphicx}
\begin{document}

\title{Discrete Spinning Tops -- Difference equations for Euler, Lagrange, and Kowalevski tops}
\author{Kiyoshi \textsc{Sogo}
\thanks{EMail: sogo@icfd.co.jp}
}
\date{}

\maketitle

\begin{center}
Institute of Computational Fluid Dynamics, 1-16-5, Haramachi, Meguroku, 
Tokyo, 152-0011, Japan
\end{center}
\abstract{
Several methods of time discretization are examined for integrable rigid body models, 
such as Euler, Lagrange, and Kowalevski tops. 
Problems of Lax-Moser pairs, conservation laws, and explicit solver algorithms are discussed. 
New discretization method is proposed for Kowalevski top, which have properties $\boldsymbol{\gamma}^2=1$, 
and the Kowalevski integral $|\xi|^2=\text{const.}$ satisfied exactly. Numerical tests are done successfully.
}
%
%

\section{Introduction}
\setcounter{equation}{0}
\subsection{Euler-Poisson equation}

Motions of general top under gravity are described by the Euler-Poisson equations \cite{Audin}
\begin{align}
\dot{\boldsymbol{m}}=\boldsymbol{m}\times\boldsymbol{\omega} + \boldsymbol{\gamma}\times\boldsymbol{g},\qquad 
\dot{\boldsymbol{\gamma}}=\boldsymbol{\gamma}\times\boldsymbol{\omega},
\label{EulerPoisson}
\end{align}
where the dot implies time differentiation $d/dt$, and $\boldsymbol{m}$ angular momentum, $\boldsymbol{\omega}$ 
angular velocity, which are related to each other by $\boldsymbol{m}=\text{diag}(A, B, C)\ \boldsymbol{\omega}$, that is,    
$m_1=A\omega_1,\ m_2=B\omega_2,\ m_3=C\omega_3$ with moments of inertia $A,\ B,\ C$, 
and the base vector $\boldsymbol{\gamma}=\boldsymbol{e}_z$, and $\boldsymbol{g}=mg(x_0, y_0, z_0)$ coordinate of 
gravitational center of the top. In components  they are given by
\begin{align}
A\dot{\omega}_1&=(B-C)\omega_2\omega_3+mg(\gamma_2 z_0-\gamma_3 y_0),\nonumber \\
B\dot{\omega}_2&=(C-A)\omega_3\omega_1+mg(\gamma_3 x_0-\gamma_1 z_0),
\label{Euler}\\
C\dot{\omega}_3&=(A-B)\omega_1\omega_2+mg(\gamma_1 y_0-\gamma_2 x_0),\nonumber \\
\dot{\gamma}_1&=\gamma_2\omega_3-\gamma_3\omega_2,\nonumber \\
\dot{\gamma}_2&=\gamma_3\omega_1-\gamma_1\omega_3,
\label{Poisson}\\
\dot{\gamma}_3&=\gamma_1\omega_2-\gamma_2\omega_1.\nonumber
\end{align}

Before proceeding further let us note that \eqref{EulerPoisson} have three conservation laws:
\begin{align}
1)\ \boldsymbol{\gamma}^2=1,\quad 
2)\ \boldsymbol{m}\cdot\boldsymbol{\gamma}=\text{const.},\quad
3)\ E=\frac{1}{2}\boldsymbol{m}\cdot\boldsymbol{\omega}+\boldsymbol{g}\cdot\boldsymbol{\gamma}=\text{const.},
\label{constant}
\end{align}
which are derived easily from \eqref{EulerPoisson}. 
If there exists the fourth integral, \eqref{EulerPoisson} becomes completely integrable. 
It is known that only three cases are completely integrable. \cite{Audin} 
\begin{align}
&(1)\quad \text{Euler:}\ (x_0, y_0, z_0)=0,\quad \boldsymbol{m}^2=\text{const.},
\\
&(2)\quad \text{Lagrange:}\ A=B,\ x_0=y_0=0,\quad \omega_3=\text{const.},
\\
&(3)\quad \text{Kowalewski:}\ A=B=2C,\ y_0=z_0=0,\quad k^2=|\omega^2-c_0\gamma|^2=\text{const.}
\end{align} 
with $\omega=\omega_1+i\omega_2,\ \gamma=\gamma_1+i\gamma_2,\ c_0=mgx_0$.  
We discuss on the problem of time discretization of these cases in this paper. 

Now let us rewrite \eqref{Euler}, \eqref{Poisson} in matrix form. If we introduce $3\times 3$ anti-symmetric matrices
\begin{align}
&M=\left(\begin{array}{ccc}
0&m_3&-m_2\\
-m_3&0&m_1\\
m_2&-m_1&0\end{array}\right),\quad
\Omega=\left(\begin{array}{ccc}
0&\omega_3&-\omega_2\\
-\omega_3&0&\omega_1\\
\omega_2&-\omega_1&0
\end{array}\right),\\
&\Gamma=\left(\begin{array}{ccc}
0&\gamma_3&-\gamma_2\\
-\gamma_3&0&\gamma_1\\
\gamma_2&-\gamma_1&0\end{array}\right),\quad
G=mg\left(\begin{array}{ccc}
0&z_0&-y_0\\
-z_0&0&x_0\\
y_0&-x_0&0
\end{array}\right),
\end{align}
the Euler-Poisson equations \eqref{Euler}, \eqref{Poisson} are rewritten in a matrix relation by 
\begin{align}
&\dot{M}=[\Omega, M]+[G,\ \Gamma] \quad\Longleftrightarrow\quad 
\dot{\boldsymbol{m}}=\boldsymbol{m}\times \boldsymbol{\omega}+\boldsymbol{\gamma}\times\boldsymbol{g}, \\
&\dot{\Gamma}=[\Omega,\ \Gamma]\quad\Longleftrightarrow\quad 
\dot{\boldsymbol{\gamma}}=\boldsymbol{\gamma}\times\boldsymbol{\omega},
\end{align}
where the symbol $[A,\ B]=AB-BA$ is a matrix commutator. 
We can rewrite further these two equations in a combined form \cite{Audin}
\begin{align}
\frac{d}{dt}\left(\Gamma+\varepsilon M\right)=[ \Omega+\varepsilon G,\ \Gamma+\varepsilon M],
\label{Audin}
\end{align}
where $\varepsilon$ is a Grassmann symbol $\varepsilon^2=0$. Noting such $\varepsilon$ is realized by 
$2\times 2$ matrix
\begin{align}
\varepsilon=\left(\begin{array}{cc}
0&1\\
0&0
\end{array}\right),
\end{align}
it is amusing to remark that 
\eqref{Audin} can be rewritten in a kind of Lax-Moser form
\begin{align}
\frac{d\mathcal{L}}{dt}=[\mathcal{M},\ \mathcal{L}],\qquad 
\mathcal{L}= \left(
\begin{array}{cc}
\Gamma&M\\
0&\Gamma
\end{array}\right),\quad
\mathcal{M}=
\left(\begin{array}{cc}
\Omega&G\\
0&\Omega
\end{array}\right),
\end{align}
however this {\it should not} be called Lax-Moser pair, since general Euler-Poisson 
equations are {\it not} integrable at all. 
 
 \subsection{Discrete Euler Top}
 
Let us explain our problem of time discretization by taking the Euler top as the simplest example. 
Equations of motion for the Euler top are, which is the special case of $\boldsymbol{g}=0$ in \eqref{EulerPoisson},
\begin{align}
A\dot{\omega}_1=(B-C)\omega_2\omega_3,\quad
B\dot{\omega}_2=(C-A)\omega_3\omega_1,\quad
C\dot{\omega}_3=(A-B)\omega_1\omega_2,\nonumber 
\end{align}
which are discretized many years ago by Hirota-Kimura \cite{HK} into
\begin{align}
\omega_1^{n+1}-\omega_1^n=\frac{h(B-C)}{2A}\left(\omega_2^{n+1}\omega_3^n+\omega_2^{n}\omega_3^{n+1}\right),
\nonumber \\
\omega_2^{n+1}-\omega_2^n=\frac{h(C-A)}{2B}\left(\omega_3^{n+1}\omega_1^n+\omega_3^{n}\omega_1^{n+1}\right),
\\
\omega_3^{n+1}-\omega_3^n=\frac{h(A-B)}{2C}\left(\omega_1^{n+1}\omega_2^n+\omega_1^{n}\omega_2^{n+1}\right),
\nonumber
\end{align}
with time increment $h$ such as $t=nh$. We call this kind of explicit solver as HK scheme. 
One of important features of HK scheme is {\it time reversal symmetry}, {\it i.e.} $h\rightarrow -h$ corresponds exchanging 
$n\leftrightarrow n+1$. 

Let us mention here the author's work \cite{SogoE} giving the discrete Lax-Moser pair for this Euler top, 
by relating it to $2\times 2$ matrix Nahm system.\cite{MW}

This HK scheme is rewritten in vector form by
\begin{align}
\boldsymbol{m}^{n+1}-\boldsymbol{m}^n=\frac{h}{2}\left(\boldsymbol{m}^{n+1}\times\boldsymbol{\omega}^n+
\boldsymbol{m}^n\times\boldsymbol{\omega}^{n+1}\right). 
\end{align}
Then it is easy to verify that $(\boldsymbol{m}^{n+1})^2\neq(\boldsymbol{m}^n)^2$, because
\begin{align}
&(\boldsymbol{m}^{n+1}+\boldsymbol{m}^n)\cdot(\boldsymbol{m}^{n+1}-\boldsymbol{m}^n)=\frac{h}{2}
(\boldsymbol{m}^{n+1}+\boldsymbol{m}^n)\cdot\left(\boldsymbol{m}^{n+1}\times\boldsymbol{\omega}^n+
\boldsymbol{m}^n\times\boldsymbol{\omega}^{n+1}\right)
\nonumber \\
&\qquad=\frac{h}{2}\left( \boldsymbol{m}^n\cdot(\boldsymbol{m}^{n+1}\times\boldsymbol{\omega}^n)+
\boldsymbol{m}^{n+1}\cdot(\boldsymbol{m}^n\times\boldsymbol{\omega}^{n+1})\right)
\nonumber \\
&\qquad=\frac{h}{2}\left(\boldsymbol{\omega}^n-\boldsymbol{\omega}^{n+1}\right)\cdot(\boldsymbol{m}^n\times\boldsymbol{m}^{n+1})
\neq 0,
\end{align}
therefore the conservation law $\boldsymbol{m}^2=\text{const.}$ is broken in HK scheme, and $O(h)$ correction term is 
necessary. Hirota-Kimura \cite{HK} discussed already such corrections, also for another broken energy conservation law 
($\boldsymbol{m}\cdot\boldsymbol{\omega}=\text{const.}$).

To preserve the moment conservation $(\boldsymbol{m}^{n+1})^2=(\boldsymbol{m}^n)^2$ exactly, we should use instead 
\begin{align}
\boldsymbol{m}^{n+1}-\boldsymbol{m}^n=\frac{h}{2}\ (\boldsymbol{m}^{n+1}+\boldsymbol{m}^n)\times\boldsymbol{\omega}^n, 
\label{EulerBS}
\end{align}
where the last $\boldsymbol{\omega}^n$ can be changed to $\boldsymbol{\omega}^{n+1}$. 
Changing to $\boldsymbol{\omega}^{n+1}$ however makes solver implicit. 
This kind of solver like \eqref{EulerBS} should be called BS scheme, since Bobenko-Suris \cite{BS}
used similar algorithm to discretize Lagrange top, which will be discussed in the next section. 
It should be noted that such BS scheme breaks time reversal symmetry. 

Finally it should be remarked here that the difference equations
\begin{align}
\boldsymbol{m}^{n+1}-\boldsymbol{m}^n=\frac{h}{4}\ (\boldsymbol{m}^{n+1}+\boldsymbol{m}^n)\times
(\boldsymbol{\omega}^{n+1}+\boldsymbol{\omega}^n), 
\end{align}
conserves both $\boldsymbol{m}^2$ and $\boldsymbol{m}\cdot\boldsymbol{\omega}$ exactly. The latter is shown by 
\begin{align}
&(\boldsymbol{\omega}^{n+1}+\boldsymbol{\omega}^n)\cdot(\boldsymbol{m}^{n+1}-\boldsymbol{m}^n)=0
\quad\Longrightarrow\quad
\boldsymbol{m}^{n+1}\cdot\boldsymbol{\omega}^{n+1}=\boldsymbol{m}^{n}\cdot\boldsymbol{\omega}^{n},
\end{align}
where identity $\boldsymbol{m}^{n+1}\cdot\boldsymbol{\omega}^n=\boldsymbol{m}^{n}\cdot\boldsymbol{\omega}^{n+1}$ is used. 
This scheme has time reversal symmetry, but this is {\it not} the explicit solver unfortunately.

\section{Discrete Lagrange top}
\setcounter{equation}{0}
\subsection{Discrete Euler-Poisson equation in HK scheme}

Our Euler-Poisson equations \eqref{Euler} \eqref{Poisson} can be discretized in HK scheme as follows, 
\begin{align}
\omega_1^{n+1}-\omega_1^n&=\frac{h(B-C)}{2A}\left(\omega_2^{n+1}\omega_3^n+\omega_2^n\omega_3^{n+1}\right)
+\frac{hmg}{2A}\left(z_0(\gamma_2^{n+1}+\gamma_2^n)-y_0(\gamma_3^{n+1}+\gamma_3^n)\right),
\nonumber \\
\omega_2^{n+1}-\omega_2^n&=\frac{h(C-A)}{2B}\left(\omega_3^{n+1}\omega_1^n+\omega_3^n\omega_1^{n+1}\right)
+\frac{hmg}{2B}\left(x_0(\gamma_3^{n+1}+\gamma_3^n)-z_0(\gamma_1^{n+1}+\gamma_1^n)\right),
\nonumber \\
\omega_3^{n+1}-\omega_3^n&=\frac{h(A-B)}{2C}\left(\omega_1^{n+1}\omega_2^n+\omega_1^n\omega_2^{n+1}\right)
+\frac{hmg}{2C}\left(y_0(\gamma_1^{n+1}+\gamma_1^n)-x_0(\gamma_2^{n+1}+\gamma_2^n)\right),
\nonumber \\
\gamma_1^{n+1}-\gamma_1^n&=\frac{h}{2}\left(\gamma_2^{n+1}\omega_3^n+\gamma_2^n\omega_3^{n+1}
-\gamma_3^{n+1}\omega_2^n-\gamma_3^n\omega_2^{n+1}\right),
\nonumber \\
\gamma_2^{n+1}-\gamma_2^n&=\frac{h}{2}\left(\gamma_3^{n+1}\omega_1^n+\gamma_3^n\omega_1^{n+1}
-\gamma_1^{n+1}\omega_3^n-\gamma_1^n\omega_3^{n+1}\right),
\nonumber \\
\gamma_3^{n+1}-\gamma_3^n&=\frac{h}{2}\left(\gamma_1^{n+1}\omega_2^n+\gamma_1^n\omega_2^{n+1}
-\gamma_2^{n+1}\omega_1^n-\gamma_1^n\omega_2^{n+1}\right),
\label{dEulerPoisson}
\end{align}
which is an explicit solver with time reversal symmetry. 
In fact, such scheme gives $(\boldsymbol{\omega}^{n+1}, \boldsymbol{\gamma}^{n+1})$ from 
$(\boldsymbol{\omega}^{n}, \boldsymbol{\gamma}^{n})$ all six components in one step by linear algebra. 
It should be noted however that three conservation laws \eqref{constant} are all broken  
unfortunately. But such violations are in practice very small fortunately. 

\subsection{Discrete Lagrange Top in HK scheme}


Kimura-Hirota \cite{KH} applied this HK scheme to the case of Lagrange top, which is $A=B,\ x_0=y_0=0$. 
Their algorithm is \eqref{dEulerPoisson} with such special parameters. 
They also gave discussions about correction terms for conservation laws. 
Here it should be remarked further that the same HK scheme \eqref{dEulerPoisson} can also be applied to discrete Kowalevski top 
by setting $A=B=2C,\ y_0=z_0=0$. Application to Kowalevski top will be shown in the next section. 
In conclusion, the HK scheme difference equations have advantage to be an explicit solver, but have disadvantage 
that obtained results do not satisfy conservation laws exactly, while it is allowable practically.

\subsection{Discrete Lagrange top in BS scheme}

On the other hand, Bobenko-Suris \cite{BS} considered Lagrange top not in the moving frame but in the inertial 
coordinate frame, which makes the equations simpler. 
The author \cite{SogoL} derived such equations of motion,  gave conservation laws and obtained 
exact solutions both in continuous and discrete cases. Let us summarize the results briefly. 

Difference equations for discrete Lagrange top are, in dimensionless form,
\begin{align}
&\boldsymbol{m}^{n+1}-\boldsymbol{m}^n=h\ \boldsymbol{p}\times\boldsymbol{a}^n,
\label{L1}\\
&\boldsymbol{a}^{n+1}-\boldsymbol{a}^n=\frac{h}{2}\ \boldsymbol{m}^{n+1}\times
\left(\boldsymbol{a}^{n+1}+\boldsymbol{a}^n\right),
\label{L2}
\end{align}
where $\boldsymbol{m}$ is the angular momentum, $\boldsymbol{a}$ is the base vector $\boldsymbol{e}_3$ which moves 
in inertial frame, and $\boldsymbol{p}$ is $\boldsymbol{e}_z$ which is a constant vector.
We can recognize easily that these difference equations are two steps explicit solver. 

Now conservation laws are
\begin{align}
&1)\ (\boldsymbol{a}^n)^2=1,\\
&2)\ \boldsymbol{m}^n\cdot\boldsymbol{p}=\text{const.},\\
&3)\ \boldsymbol{m}^n\cdot\boldsymbol{a}^n=\text{const.},\\
&4)\ E=\frac{1}{2}(\boldsymbol{m}^n)^2+\boldsymbol{a}^n\cdot\boldsymbol{p}+
\frac{h}{2}(\boldsymbol{a}^n\times\boldsymbol{m}^n)\cdot\boldsymbol{p}=\text{const.}
\end{align}
The first law is shown from \eqref{L2} by the same logic of BS scheme before. 
Take a glance at the $O(h)$ term in $E$, the energy conservation law. 
Other proofs including this are given in the paper.\cite{SogoL} Exact solutions in terms of Jacobi's elliptic functions are also derived, 
but are all omitted here.  

\section{Discrete Kowalevski top}
\setcounter{equation}{0}
\subsection{Kowalevski Top and Conservation Laws}
For the sake of simplicity, we set $A=B=2,\ C=1,\ y_0=z_0=0$ and $mgx_0=c_0$ without loss of generality. 
We keep the parameter $c_0$ for a while, although we set $c_0=1$  in numerical simulations later. 
Then we have
\begin{align}
&\dot{\omega}_1=\frac{1}{2}\omega_2\omega_3,\quad
\dot{\omega}_2=-\frac{1}{2}\omega_3\omega_1+\frac{c_0}{2}\gamma_3,\quad
\dot{\omega}_3=-c_0\gamma_2, 
\label{K1}\\
&\dot{\gamma}_1=\gamma_2\omega_3-\gamma_3\omega_2,\quad
\dot{\gamma}_2=\gamma_3\omega_1-\gamma_1\omega_3,\quad
\dot{\gamma}_3=\gamma_1\omega_2-\gamma_2\omega_1.
\label{K2}
\end{align}
If we introduce $\omega=\omega_1+i\omega_2,\ \overline{\omega}=\omega_1-i\omega_2$, and 
$\gamma=\gamma_1+i\gamma_2,\ \overline{\gamma}=\gamma_1-i\gamma_2$, we have
\begin{align}
&\frac{d\omega}{dt}=-\frac{i}{2}\left(\omega_3\omega-c_0\gamma_3\right),\quad
\frac{d\overline{\omega}}{dt}=+\frac{i}{2}\left(\omega_3\overline{\omega}-c_0\gamma_3\right),\quad
\frac{d\omega_3}{dt}=-c_0\gamma_2,\\
&\frac{d\gamma}{dt}=-i\left(\omega_3\gamma-\omega\gamma_3\right),\quad 
\frac{d\overline{\gamma}}{dt}=+i\left(\omega_3\overline{\gamma}-\overline{\omega}\gamma_3\right),\quad 
\frac{d\gamma_3}{dt}=\frac{i}{2}\left(\overline{\omega}\gamma-\omega\overline{\gamma}\right),
\end{align}
from which we have especially
\begin{align}
&\frac{d}{dt}\omega^2=-i\omega_3\omega^2+ic_0\gamma_3\omega,\quad
c_0\frac{d}{dt}\gamma=-ic_0\omega_3\gamma+ic_0\gamma_3\omega
\nonumber \\
&\Longrightarrow\quad
\frac{d}{dt}\left(\omega^2-c_0\gamma\right)=-i\omega_3\left(\omega^2-c_0\gamma\right),
\label{Kxi}
\end{align}
which implies the existence of the fourth integral, Kowalevski's conservation law
\begin{align}
\left|\omega^2-c_0\gamma\right|^2=\xi\overline{\xi}=\text{const.}=k^2,
\end{align}
where we set $\xi=\omega^2-c_0\gamma,\ \overline{\xi}=\overline{\omega}^2-c_0\overline{\gamma}$ according to Kowalevski 
\cite{Kowalevski}, while her $c_0$ has opposite sign due to her choosing opposite sign of potential energy, 
which seems to be a strange tradition of her age common to {\it e.g.} C.G. Jacobi, C. Neumann, and E. Cartan.  

In summary, conservation laws for Kowalevski top are
\begin{align}
&1)\quad \boldsymbol{\gamma}^2=\gamma_1^2+\gamma_2^2+\gamma_3^2=1,
\nonumber \\
&2)\quad 2\ell=2(\omega_1\gamma_1+\omega_2\gamma_2)+\omega_3\gamma_3=\text{const.},
\nonumber \\
&3)\quad E=\omega_1^2+\omega_2^2+\frac{1}{2}\omega_3^2+c_0\gamma_1=\text{const.},
\nonumber \\
&4)\quad k^2=\xi\overline{\xi}=|\omega^2-c_0\gamma|^2=\text{const.},
\nonumber
\end{align}
which are all four conserved quantities. 

\subsection{Discrete, Explicit Schemes for Kowalevski Top}

\subsubsection{ Kowalevski top in HK scheme}

Figure 1 shows calculations in HK scheme mentioned before. 
Doing numerical simulations we set $c_0=1$ and put
\begin{align}
\omega_1=2,\quad \omega_2=\omega_3=0,\quad \gamma_1=\sqrt{1-\gamma_3^2},\quad \gamma_2=0,\quad \gamma_3=0.001, 
\end{align}
as the initial condition, which are the same values used by Yoshida. \cite{Yoshida} 
Calculations are made for $h=0.001$ over $N=50000$ steps, by using GSL (Gnu Scientific Library) package. 
Results in Fig.1 are three components of $\boldsymbol{\gamma}$ (left) and $\boldsymbol{\omega}$ (right). 
Our results show almost the same behaviors as Yoshida's, whose computation scheme was not described in the paper, 
might be however a pre-version of his famous symplectic algorithm. \cite{YoshidaYS4} 
Kowalevski's constant $k^2$ changes between $9.000003$ and $9.000011$ osscilatory.

\begin{figure}[h]
\centering
\includegraphics[width=4.5cm]{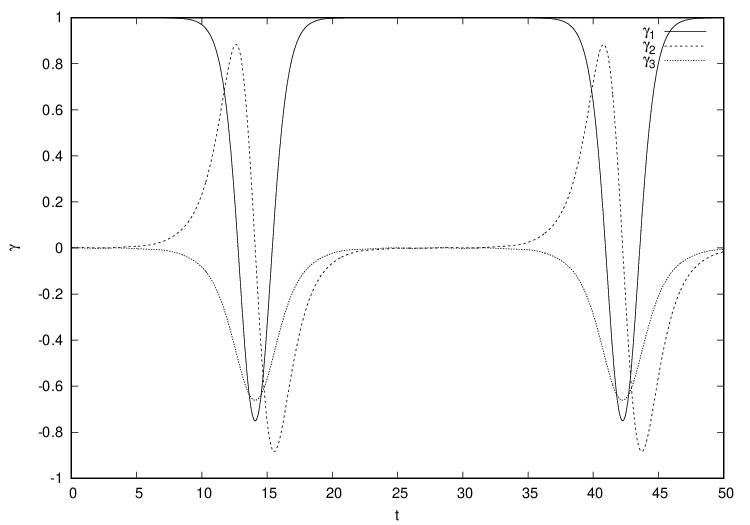}
\includegraphics[width=4.5cm]{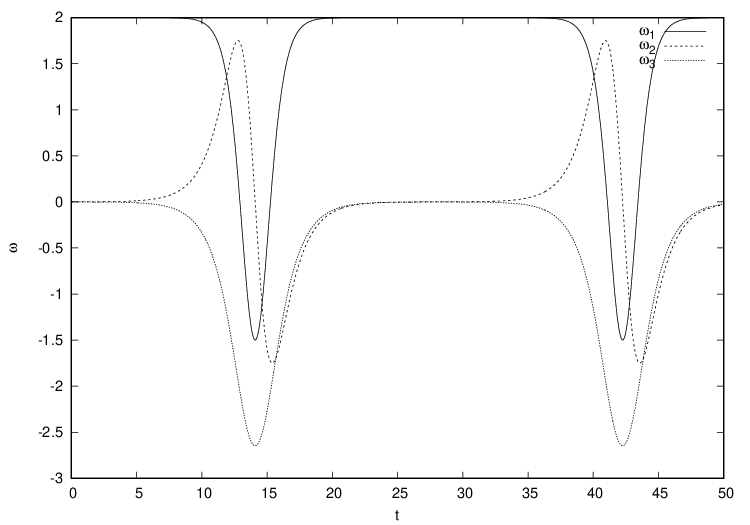}
\caption{Kowalevski top $\boldsymbol{\gamma}$ (left) and $\boldsymbol{\omega}$ (right) by HK scheme}
\end{figure}

\subsubsection{Schemes to hold conservation laws}

Let us propose new discretization methods which satisfy at least the following two conservation laws 
\begin{align}
&1)\quad \boldsymbol{\gamma}^2=1,\ \text{and}
\\
&4)\quad \xi\overline{\xi}=k^2,\qquad 
\xi=\omega^2-c_0\gamma,\quad \overline{\xi}=\overline{\omega}^2-c_0\overline{\gamma},
\end{align}
exactly. 

\noindent
(1)\ Methods to keep $\boldsymbol{\gamma}^2=1$

For this purpose, we have at least three methods, one is the BS scheme
\begin{align}
(\text{A})\quad \boldsymbol{\gamma}^{n+1}-\boldsymbol{\gamma}^n=\frac{h}{2}
\left(\boldsymbol{\gamma}^{n+1}+\boldsymbol{\gamma}^n\right)\times\boldsymbol{\omega}^n.
\end{align}
The next is to use variable transformation by introducing a complex variable $z$ 
\begin{align}
(\text{B})\quad &\gamma=\frac{2z}{1+z\overline{z}},\quad \overline{\gamma}=\frac{2\overline{z}}{1+z\overline{z}},\quad
\gamma_3=\frac{1-z\overline{z}}{1+z\overline{z}}
\\
&\Longleftrightarrow\quad z=\frac{\gamma_1+i\gamma_2}{1+\gamma_3},\quad 
\overline{z}=\frac{\gamma_1-i\gamma_2}{1+\gamma_3},\quad
z\overline{z}=\frac{1-\gamma_3}{1+\gamma_3},
\end{align}
which is a stereo-graphic transformation. Condition $\boldsymbol{\gamma}^2=1$ is satisfied automatically. 
In terms of $z,\ \overline{z}$, \eqref{K2} is rewritten as
\begin{align}
\dot{z}=\frac{i}{2}\left(\omega-2\omega_3z-\overline{\omega}z^2\right),\qquad
\dot{\overline{z}}=-\frac{i}{2}\left(\overline{\omega}-2\omega_3\overline{z}-\omega\overline{z}^2\right),
\end{align}
which are easily time discretized by {\it e.g.} Euler method.  

The last method is to interpret \eqref{K2} as an instantaneous 
rotation by $\boldsymbol{\omega}$ with time lapse $h$
\begin{align}
(\text{C})\quad \boldsymbol{\gamma}^{n+1}=R(\boldsymbol{\theta})\ \boldsymbol{\gamma}^n,\qquad 
\boldsymbol{\theta}=h \boldsymbol{\omega}^n
\end{align}
where $3\times 3$ rotation matrix $R$ is given by
\begin{align}
&R(\boldsymbol{\theta})=\left(\begin{array}{ccc}
n_1^2+(1-n_1^2)c&n_1n_2(1-c)+n_3s&n_1n_3(1-c)-n_2s\\
n_2n_1(1-c)-n_3s&n_2^2+(1-n_2^2)c&n_2n_3(1-c)+n_1s\\
n_3n_1(1-c)+n_2s&n_3n_2(1-c)-n_1s&n_3^2+(1-n_3^2)c
\end{array}\right),
\\
&\qquad \boldsymbol{n}=\boldsymbol{\theta}/\theta,\quad \theta=|\boldsymbol{\theta}|,\quad c=\cos\theta,\quad s=\sin\theta.
\end{align}
These three methods break time reversal symmetry unfortunately, and will be tested later.

\noindent
(2)\ Method to keep $\xi\overline{\xi}=k^2$

Our equation \eqref{Kxi} can be integrated for small $h$ by trapezoidal rule as follows
\begin{align}
&\frac{d}{dt}\log\xi=-i\omega_3
\quad\Longrightarrow\quad
\log\left(\frac{\xi^{n+1}}{\xi^n}\right)=-\frac{ih}{2}\left(\omega_3^{n+1}+\omega_3^n\right)
\nonumber \\
&\Longrightarrow\quad \xi^{n+1}=e^{-i\chi}\ \xi^n,\quad \chi=\frac{h}{2}\left(\omega_3^{n+1}+\omega_3^n\right)
\end{align}
that is, 
\begin{align}
\left(\omega^2-c_0\gamma\right)^{n+1}=e^{-i\chi}\left(\omega^2-c_0\gamma\right)^n.
\end{align}
Since $\gamma^n,\ \gamma^{n+1},\ \omega^n$ and $\omega_3^n,\ \omega_3^{n+1}$ (therefore $\chi$ also) 
are already known, this is an equation to be solved for the unknown $\omega^{n+1}$, such as
\begin{align}
\left(\omega_1^{n+1}+i\omega_2^{n+1}\right)^2=e^{-i\chi}\left[\left(\omega_1^n+i\omega_2^n\right)^2-c_0\gamma^n\right]
+c_0\gamma^{n+1},
\end{align}
where the right hand side is known and computable. This type of equation $w^2=z$ is similar to  
{\it Bohlin transformation} \cite{Arnold}
which relates Kepler with Hooke (harmonic oscillator), both 2 dimensional motion. 
Let us call therefore our method also Bohlin's method. 
It should be noted this Bohlin's method has time reversal symmetry. 

Solution of $w^2=z$ is one of $w=\pm \sqrt{z}$, or
\begin{align}
w=r\ e^{i\phi},\quad z=R\ e^{i\Phi}
\quad\Longrightarrow\quad
r=\sqrt{R},\quad \phi=\frac{\Phi}{2}\quad \text{mod}\ 2\pi,
\end{align}
where we must carefully choose the correct angle $\phi$. For such purpose, we solve once
\begin{align}
\frac{d\omega}{dt}=-\frac{i}{2}\left(\omega_3\omega-c_0\gamma_3\right)
\quad\Longrightarrow\quad
\omega^{n+1}=\omega^n-\frac{ih}{2}\left(\omega_3^n\omega^n-c_0\gamma_3^n\right),
\label{direct}
\end{align}
only to find the correct branch of phase $\phi$ of $\omega^{n+1}=r\ e^{i\phi}$. 
To state precisely, it is sufficient to compare two arguments of complex $\omega^{n+1}$'s, one from Bohlin's method, 
the other from \eqref{direct}. 
We need only the sign (plus or minus) of argument, which is obtained by {\it e.g.} atan2 function of C Language, 
and change sign of the former (Bohlin's) if two are different.

In summary our algorithm to find $(\boldsymbol{\gamma}^{n+1},\ \boldsymbol{\omega}^{n+1})$ from 
$(\boldsymbol{\gamma}^{n},\ \boldsymbol{\omega}^{n})$ in three steps is as follows:
\begin{align}
&(1)\quad \text{Find } \boldsymbol{\gamma}^{n+1} \text{ by one of methods\ } (\text{A}),\ (\text{B})\ \text{or}\ (\text{C}).
\nonumber \\
&(2)\quad \text{Set } \omega_3^{n+1}=\omega_3^n-\frac{h}{2}c_0\left(\gamma_2^{n+1}+\gamma_2^n\right).
\nonumber \\
&(3)\quad \text{Find } (\omega_1^{n+1},\ \omega_2^{n+1})\ \text{by Bohlin's method}.
\nonumber
\end{align}

Figure 2 shows the results of Bohlin's method with (A), taking the same initial conditions as Fig.1. 
Since differences among methods (A) to (C) are indistinguishable, we omit their figures. 
In other words our three methods (A) to (C) keeping $\boldsymbol{\gamma}^2=1$ have almost the same quality. 

\begin{figure}[h]
\centering
\includegraphics[width=4.5cm]{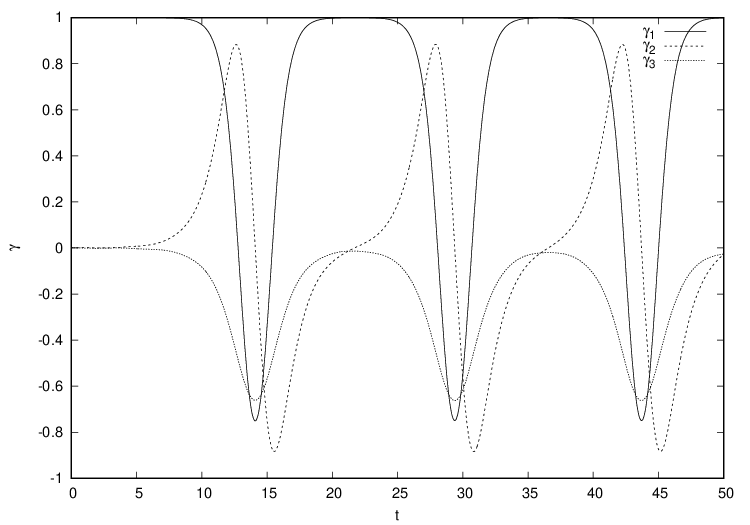}
\includegraphics[width=4.5cm]{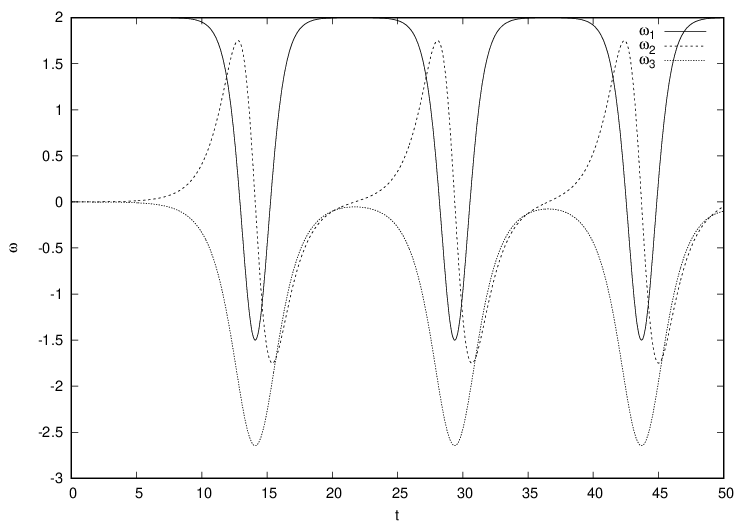}
\caption{Kowalevski top $\boldsymbol{\gamma}$ (left) and $\boldsymbol{\omega}$ (right) by Bohlin with (A) method}
\end{figure}

Obvious but strange difference found by comparing Fig.2 with Fig.1 is the magnitudes of period, even though 
the same initial conditions and parameters are used. 
This implies 
\begin{center}
{\it the time reversal symmetry is extremely important}
\end{center}
for the dynamical behaviors of Kowalevski top. 

It is also recognized from Fig.3, whose left shows conserved quantities 2) $2\ell=\text{const.}$ 
and 3) $E=\text{const.}$ by Bohlin with (A) method.  
We observe that both quantities seem to be constant indeed. However 
Fig.3 right shows the energy $E$ actually makes a slow and tiny increase periodically, look at the vertical scale carefully. 
Behavior of $2\ell$ is almost the same. These behaviors are surely consequences of time reversal asymmetry. 

\begin{figure}[h]
\centering
\includegraphics[width=4.5cm]{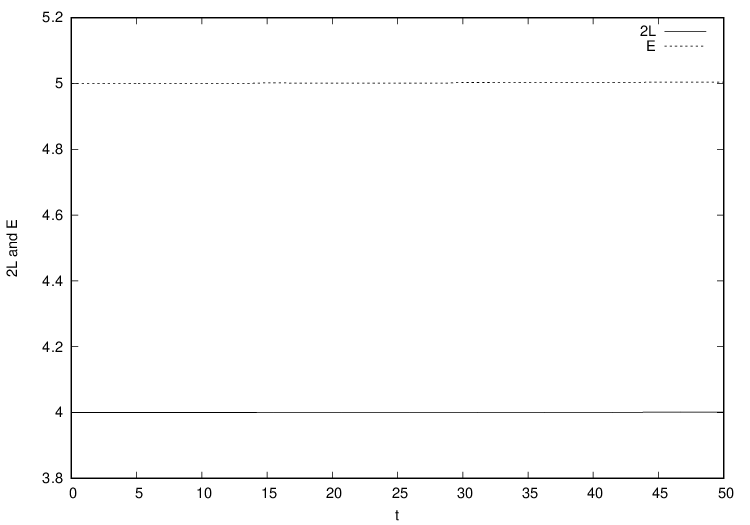}
\includegraphics[width=4.5cm]{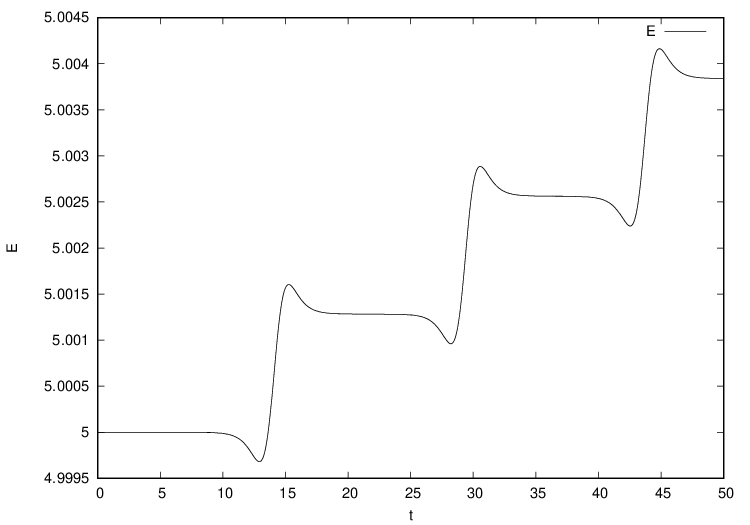}
\caption{Kowalevski top $2\ell$ and $E$ by Bohlin with (A) method}
\end{figure}

\subsubsection{Hybrid of two schemes}

In this way, we have arrived finally at a hybrid scheme. Our proposal is as follows. 
\begin{align}
&(1)\quad \text{Find}\ (\boldsymbol{\gamma}^{n+1},\ \boldsymbol{\omega}^{n+1})\ \text{ by HK scheme once.}
\nonumber \\
&(2)\quad \text{Solve}\ \boldsymbol{\gamma}^{n+1}\ \text{again by modified BS scheme}
\nonumber \\
&\qquad\qquad \boldsymbol{\gamma}^{n+1}-\boldsymbol{\gamma}^n=\frac{h}{4}\left(\boldsymbol{\gamma}^{n+1}+
\boldsymbol{\gamma}^n\right)
\times\left(\boldsymbol{\omega}^{n+1}+\boldsymbol{\omega}^n\right).
\\
&(3)\quad \text{Find}\ (\omega_1^{n+1},\ \omega_2^{n+1})\ \text{again by Bohlin's method.}
\nonumber
\end{align}
This scheme keeps $\boldsymbol{\gamma}^2=1$ and $k^2=|\omega-c_0\gamma|^2=\text{const.}$, 
and also has time reversal symmetry.  Obtained Figure 4 looks almost the same as Figure 1, 
except that $\boldsymbol{\gamma}^2=1$ and $k^2=|\omega^2-c_0\gamma|^2=\text{const.}$ satisfied exactly. 

\begin{figure}[h]
\centering
\includegraphics[width=4.5cm]{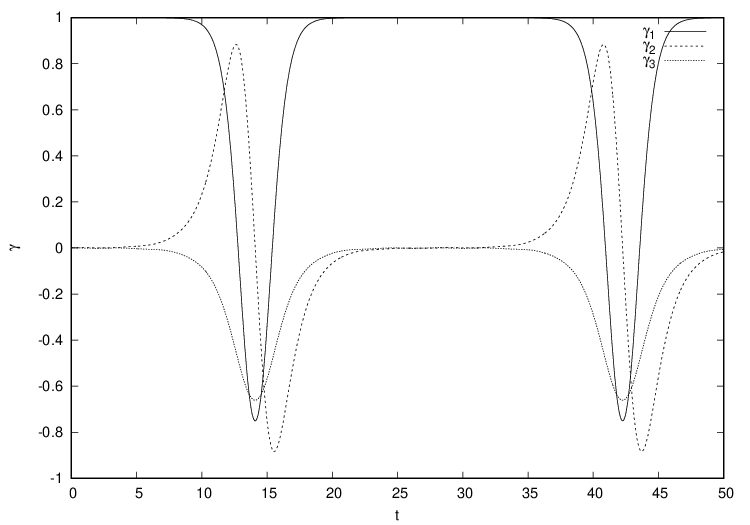}
\includegraphics[width=4.5cm]{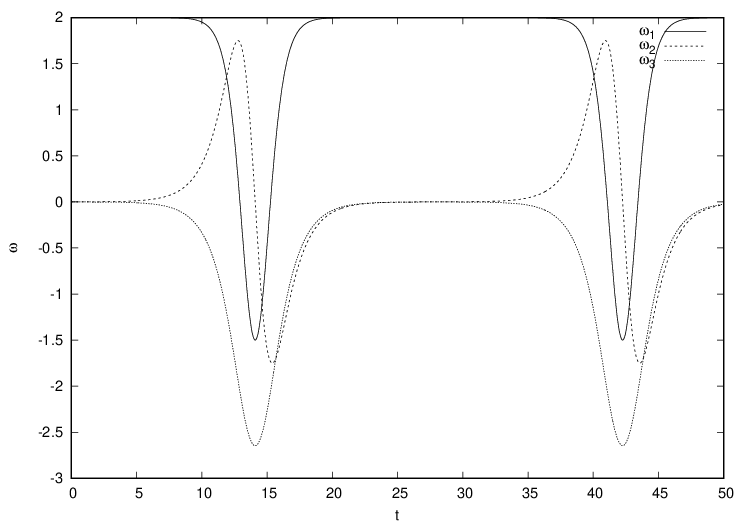}
\caption{Kowalevski top $\boldsymbol{\gamma}$ (left) and $\boldsymbol{\omega}$ (right) by hybrid method}
\end{figure}

\section{Summary and remarks}
\setcounter{equation}{0}
General Euler-Poisson equations can be discretized by Hirota-Kimura method, which can be applied even to Euler, 
Lagrange, and Kowalevski tops. One of its defects however is to break conservation laws, although discrepancies 
are very small. To remedy such defects several methods are examined on account of accuracy and efficiency. 
In the course it is found that time reversal symmetry is very important. 
For Kowalevski top, an hybrid scheme of Hirota-Kimura method and Bohlin's method, which is newly found and named, 
is proposed, and is numerically tested successfully. 

Present discretization scheme is, in a sense, merely an approximation to the continuous-time equation. 
On the contrary, it is desirable to find the discrete-time equation which is also completely integrable by itself. 
Many years ago, Haine-Horozov \cite{HaineHorozov} wrote that Kowalevski top is equivalent to Clebsch (or Neumann) system, 
which is also integrable. 
However their equivalence needs unfortunately reconsiderations, since their relationships given are {\it not} 
real to real variables correspondence. 
If any resolution to such problem is completed, the {\it integrable and discrete} Kowalevski system 
will be constructed finally.

\end{document}